\documentstyle[12pt]{article}
\begin{document}
\baselineskip 5mm

\newcommand{\ov}{\overline}
\newcommand{\r}{$^{[??]}$}

\newcommand{\bops}{\mathop{\textstyle\otimes}}

\title{ The Fuzzy K\"ahler Coset Space  \\ 
 by the Fedosov Formalism }

\author{Shogo Aoyama\thanks{e-mail: spsaoya@ipc.shizuoka.ac.jp}\ \ \ \ and 
\ \ Takahiro Masuda\thanks{e-mail: stmasud@ipc.shizuoka.ac.jp} \\
      \\
       Department of Physics \\
              Shizuoka University \\
                Ohya 836, Shizuoka  \\
                 Japan}
                 
\maketitle

\begin{abstract}
We discuss  deformation quantization of the K\"ahler coset space by using the Fedosov formalism. We show that the Killing potentials of the K\"ahler coset space satisfy the fuzzy algebrae, when the coset space is irreducible. 
\end{abstract}

\newpage

Quantum field theory on a non-commutative space has recently raised much interest mainly due to the connection with the string theory\cite{1}. There appears non-commutativity of the flat target space via a constant $B$-field. The ordinary product of fields in the effective theory is replaced by the non-commutative one, called the Moyal product. A lot of works have been done on the physics caused by this non-commutativity of the flat target space\cite{1'}. It is of obvious interest to generalize the studies to the string theory with a non-constant $B$-field and a curved target space\cite{1''}.  Then the Moyal product should be replaced by a more general non-commutative one, called the $\star$ product.

Generalization of the Moyal product has a long history\cite{2} in the study of deformation quantization on symplectic manifolds or more generally Poisson manifolds. Several definitions of the $\star$ product exist in the literature. However it has been discussed in the physical context in few references. The work by Cattaneo and Felder\cite{3} which gave a quantum field theoretical interpretation to  Kontsevich's $\star$ product\cite{4} is one of them. The recent work by Kishimoto\cite{5} was also done from motives for physical application. Namely he explicitly constructed the $\star$ product on the fuzzy sphere $S^2$ by using the Fedosov formalism for deformation quantization\cite{6}. With  the explicit form of the $\star$ product  it was shown   that there exist the fuzzy sphere algebrae\cite{7}, which may be useful for studying non-commutative non-linear $\sigma$-models.
 
 In this paper we generalize his arguments to the fuzzy K\"ahler coset space $G/H$. 
The key observation is that the ordinary K\"ahler coset space has a set of the Killing potentials $M^A(z,\ov z)$ satisfying 
\begin{eqnarray}
\sum_{A=1}^{dim\ G} M^A(z,\ov z)M^A(z,\ov z) \ =\  R.  \nonumber
\end{eqnarray}
Here $R$ is  the Riemann scalar curvature of $G/H$, which is a constant.
We study the $\star$ product by the Fedosov formalism on the K\"ahler coset space 
$G/H$ and show that the Killing potentials satisfy the fuzzy algebrae 
\begin{eqnarray}
M^A(z,\ov z)\star M^B(z,\ov z) &-& M^B(z,\ov z)\star M^A(z,\ov z)  \nonumber \\
&=& -i(c_1 \hbar + c_3\hbar^3 + c_5\hbar^5 + \cdots)\sum_{A=1}^{dim\ G}f^{ABC} M^C(z,\ov z),  \label{eq 0} \\
\sum_{A=1}^{dim\ G} M^A(z,\ov z)\star M^A(z,\ov z) &=& R  + c_2\hbar^2 + c_4\hbar^4 + \cdots\ ,  \label{eq 00}
\end{eqnarray}
when the coset space is irreducible.
The coefficients $c_1,c_2,c_3, \cdots $ are numerical constants. As a demonstration, they will be calculated to order of $\hbar^4$ for  the case of $CP^1$.

\hspace{2cm}

We start with a brief review on the Fedosov formalism for deformation quantization\cite{6}.
We consider a real $2N_0$-dimensional Riemann manifold $\cal M$ with local coordinates  $x^a =(x^1,x^2,\cdots\cdots,x^{2N_0})$. The line element of the manifold is given by 
\begin{eqnarray}
  ds^2 = g_{ij}dx^i dx^j.      \label{eq 1}
\end{eqnarray}
Suppose that it is endowed with a symplectic structure 
 given by a non-degenerate 2-form 
\begin{eqnarray}
\omega = {1\over 2}\omega_{ij}dx^i \wedge dx^j.  \label{eq 2}
\end{eqnarray}
We  require 
$\omega_{ij}$ to be covariantly constant:
\begin{eqnarray}
\omega_{ij;k} = 0.  \label{eq 3}
\end{eqnarray}
Then (\ref{eq 2}) is closed, $d\omega = 0$. $\omega_{ij}$ is inverted by $\omega^{ij}$:
\begin{eqnarray}
\omega_{ik}\omega^{kj} = \omega^{jk}\omega_{ki} = \delta^j_i.   \label{eq 4}
\end{eqnarray}
For a differential $q-$form on $\cal M$ 
\begin{eqnarray}
a_0 = {1\over q!}a(x)_{j_1j_2\cdots j_q}\theta^{j_1}\theta^{j_2}\cdots \theta^{j_q},
\quad\quad\quad \theta^j \equiv dx^j,  \nonumber
\end{eqnarray}
we consider a deformed $q-$form
\footnote {As in ref. \cite{6} we may consider a more general deformed form  summing  over the rank $q$. But we here make a simplification to avoid unnecessary complications for our scope.}
\begin{eqnarray}
a = \sum^{\infty}_{p=0}{1\over p!q!}a(x)_{i_1i_2\cdots i_pj_1j_2\cdots j_q}
  y^{i_1}y^{i_2}\cdots y^{i_p}\theta^{j_1}\theta^{j_2}\cdots \theta^{j_q},
  \nonumber 
\end{eqnarray}
in which $y^i = (y^1,y^2,\cdots ,y^{2N_0})$ are deformation variables. 
The coefficient $a(x)_{i_1i_2\cdots i_pj_1j_2\cdots j_q}$ is a covariant tensor which is symmetric in $i_1,i_2,\cdots i_p$ and anti-symmetric in $j_1,j_2,\cdots j_q$.
We define various operations for the deformed differential form:
\begin{eqnarray}
 a \circ b &=& \sum_{n=0}^{\infty}{1\over n!}(-{i\hbar \over 2})^n \omega^{i_1 j_1}\omega^{i_2 j_2}
\cdots \omega^{i_n j_n}\partial_{i_1}\partial_{i_2}\cdots \partial_{i_n}a
 \partial_{j_1}\partial_{j_2}\cdots \partial_{j_n}b,  \label{eq 5} \\
\partial a &=&  \sum^{\infty}_{p=0}{1\over p!q!}\theta^l\nabla_l^x
a(x)_{i_1i_2\cdots i_pj_1j_2\cdots j_q}
  y^{i_1}y^{i_2}\cdots y^{i_p}\theta^{j_1}\theta^{j_2}\cdots \theta^{j_q},\label{eq 6} \\
\delta a &=& \theta^l {\partial \over \partial y^l}a, \label{eq 7} \\
   \delta^{-1} a &=& \sum_{p=0}^\infty {1\over p+q}y^l {\partial \over \partial \theta^l}
[{1\over p!q!}a(x)_{i_1i_2\cdots i_pj_1j_2\cdots j_q}
  y^{i_1}y^{i_2}\cdots y^{i_p}\theta^{j_1}\theta^{j_2}\cdots \theta^{j_q}].\label{eq 8}
\end{eqnarray}
In the $\circ$-product (\ref{eq 5}) $\hbar$ is a deformation parameter. To $\hbar$ and $y^i$ we assign degrees 2 and 1 respectively. Then the $\circ$ product preserves the degree.
In the Fedosov formalism the graded commutator is defined with this $\circ$ product:
\begin{eqnarray}
[a ,b] = a\circ b - (-1)^{q_a q_b}b\circ a,   \nonumber
\end{eqnarray}
in which $q_a$ and $q_b$ are the rank of the differential forms $a$ and $b$ respectively. In (\ref{eq 6}) $\nabla^x_l$ is the covariant derivative with respect to $x$. The graded differentials $\partial, \delta $ and $\delta^{-1}$ are shown to satisfy the relations
\begin{eqnarray}
\delta^2 &=&  \delta^{-2} \ =\  0,  \nonumber \\
(\delta^{-1}\delta + \delta\delta^{-1})a &=& a - a_0,  \label{eq 9}   \\
\delta\partial + \partial\delta &=& 0,  \nonumber  \\
\partial^2 a &=& {i\over \hbar}[{\cal R}  , a ],  \label{eq 10}
\end{eqnarray}
with
\begin{eqnarray}
{\cal R} = {1\over 4}R_{ijk}^{\ \ \  m}\omega_{ml}\theta^i\theta^j y^k y^l. \label{eq 11}
\end{eqnarray}
Here $R_{ijk}^{\ \ \ m}$ is the Riemann curvature. \footnote {From (\ref{eq 3}) we can show that $R_{ijk}^{\ \ \  m}\omega_{ml} = R_{ijl}^{\ \ \  m}\omega_{mk}.$}

\vspace{1cm}

We define a more general derivative by
$$
Da = \partial a - \delta a + {i\over \hbar }[r, a],
$$
with an appropriate deformed 1-form $r$. A simple calculation gives 
$$
D^2 a = {i\over \hbar}[\Omega, a],
$$
with 
$$
\Omega = -\omega + {\cal R} - \delta r + \partial r + {i\over 2\hbar}[r,r]. 
$$
If $r$ satisfies the equation 
\begin{eqnarray}
\delta r = {\cal R}  + \partial r + {i\over 2\hbar}[r,r], \label{eq 12}
\end{eqnarray}
we have $\Omega = -\omega $ so that $D^2 a = 0$ for any $a$. Eq. (\ref{eq 12}) has a unique solution $r$ obeying the condition 
\begin{eqnarray}
\delta^{-1} r = 0.  \label{eq 13}
\end{eqnarray}
Namely owing to (\ref{eq 9}) and (\ref{eq 13}), eq. (\ref{eq 12}) may be written in the form 
\begin{eqnarray}
r = r_0 + \delta^{-1} (\partial r + {i\over 2\hbar}[r,r]),\label{eq 14}
\end{eqnarray}
with $r_0 = \delta^{-1}{\cal R}$. Note that $deg\ r_0 = 3$ and $\delta^{-1}$ raises the degree at least by 1. Hence we can solve (\ref{eq 14}) by iteration, expanding $r$ in series
\begin{eqnarray}
r = r_0 + r_1 + r_2 +\cdots,\label{eq 15}
\end{eqnarray}
with $deg\ r_n = n+3$. Conversely the iterative solution satisfy (\ref{eq 13}) by the construction. 
This solution $r$ plays a particular role in the Fedosov formalism. Provided with it, we may impose the constraint on $a$, $Da = 0$ , i.e., 
\begin{eqnarray}
\delta a = \partial a + {i\over \hbar}[r,a]. \label{eq 16}
\end{eqnarray}
When $a$ is a 0-form, one can similarly show that eq. (\ref{eq 16}) has a unique solution obeying the condition 
\begin{eqnarray}
\delta^{-1} a = 0.\label{eq 17}
\end{eqnarray}
To this end write eq. (\ref{eq 16}) in the form
\begin{eqnarray}
a = a_0 + \delta^{-1}(\partial a + {i\over \hbar }[r,a]).\label{eq 18}
\end{eqnarray}
With $a_0$ as an initial condition the solution in the form
$$
a = a_0 + a_1 + a_2 + \cdots,
$$
with $deg\ a_n = n $, may be obtained by iteration. It is obvious that the iterative solution satisfies the condition (\ref{eq 17}) because $\delta^{-1} a_0 = 0$. So the constraint (\ref{eq 16}) gives a unique way to deform 
the function $a_0 (x)$.
For $a$ and $b$ thus deformed we define the $\star$ product by
\begin{eqnarray}
a_0(x)\star b_0(x) = \hspace{10cm}  \nonumber \\
 \sum_{n=0}^{\infty} {1\over n!}(-{i\hbar \over 2})^n 
  \omega^{i_1j_1}\omega^{i_2j_2}\cdots \omega^{i_nj_n}
  {\partial \over \partial y^{i_1}}{\partial \over \partial y^{i_2}}\cdots 
  {\partial \over \partial y^{i_n}}a
  {\partial \over \partial y^{j_1}}{\partial \over \partial y^{j_2}}\cdots 
  {\partial \over \partial y^{j_n}}b \Biggr|_{y = 0}.\label{eq 19}
\end{eqnarray}
According to Fedosov\cite{6} this $\star$ product satisfies the associativity. It reduces to the Moyal product when the manifold ${\cal M}$ is flat. 

\hspace{2cm}

We shall apply the Fedosov formalism in the case where  ${\cal M}$ is the K\"ahler manifold. The K\"ahler manifold has local complex coordinates $z^\alpha =(z^1,z^2,\cdots ,z^{N_0})$ and their complex conjugates. The line element (\ref{eq 1}) and the symplectic 2-form (\ref{eq 2}) respectively reduce to 
\begin{eqnarray}
ds^2 &=& g_{\alpha\ov \beta}\ dz^\alpha dz^{\ov \beta},   \nonumber \\
\omega &=& {1\over 2}\omega_{\alpha\ov \beta}\ dz^\alpha\wedge dz^{\ov \beta},\label{eq 20}
\end{eqnarray}
with 
\begin{eqnarray}
\omega_{\alpha\ov \beta} = ig_{\alpha\ov \beta}.\label{eq 21}
\end{eqnarray}
From (\ref{eq 4}) we have 
\begin{eqnarray}
\omega^{\alpha\ov \beta} = {1\over 2}\omega^{\ov \beta\alpha}= ig^{\alpha\ov \beta}.\label{eq 22}
\end{eqnarray}
In these complex coordinates the Riemann-Christophel symbols are given by 
\begin{eqnarray}
\Gamma_{\alpha\beta}^{\ \ \gamma} = g^{\gamma\ov \eta}g_{\alpha\ov\eta,\beta},
\quad\quad 
\Gamma_{\ov\alpha\ov\beta}^{\ \ \ov\gamma} = g^{\eta\ov\gamma}g_{\eta\ov\alpha,\ov\beta},     \nonumber
\end{eqnarray}
and other components are vanishing, due to $d\omega = 0$.  
The non-trivial components of the Riemann curvature take the forms
\begin{eqnarray}
R_{\alpha\ov\beta\gamma}^{\ \ \ \ \delta} = R_{\alpha\ov\beta\gamma\ov\eta}g^{\delta\ov\eta} = \Gamma_{\alpha\gamma, \ \ov\beta}^{\ \ \delta},  \nonumber
\end{eqnarray}
and satisfy the symmetry
\begin{eqnarray}
R_{\alpha\ov\beta\gamma\ov\delta} =
-R_{\ov\beta\alpha\gamma\ov\delta} =
-R_{\alpha\ov\beta\ov\delta\gamma} =
R_{\gamma\ov\beta\alpha\ov\delta} =
R_{\alpha\ov\delta\gamma\ov\beta} .   \nonumber
\end{eqnarray}
Consequently eq. (\ref{eq 11}) reads 
\begin{eqnarray}
{\cal R} &=& {1\over 2}( R_{\alpha\ov\beta\gamma}^{\ \ \ \ \eta}\omega_{\eta\ov\delta} 
+  R_{\alpha\ov\beta\ov\delta}^{\ \ \ \ \ov\eta}\omega_{\ov\eta\gamma})\theta^\alpha\theta^{\ov\beta}
y^\gamma y^{\ov\delta}  \nonumber \\
 &=& iR_{\alpha\ov\beta\gamma\ov\delta}\theta^\alpha\theta^{\ov\beta}
y^\gamma y^{\ov\delta} , \label{eq 23}
\end{eqnarray}
by (\ref{eq 21}). It is worth cheking that the last formula indeed satisfies (\ref{eq 10}) with (\ref{eq 22}).

\hspace{2cm}

When the K\"ahler manifold is a coset space $G/H$, the Fedosov formalism is further simplified. Here we summarize the useful properties of the K\"ahler coset space for our  discussion. There exists a set of holomorphic Killing vectors
\begin{eqnarray}
R^{Aa} = (R^{1a},R^{2a},\cdots \cdots, R^{Da}), \nonumber
\end{eqnarray}
with $D = {\rm dim}\ G$, which represents the isometry $G$:
\begin{eqnarray} 
 R^{Ab} R^{Ba}_{\ \  ,b} - R^{B\ b} R^{Aa}_{\ \ ,b} 
  \ \ = \ \ f^{ABC}R^{Ca}.    \label{eq 24}
\end{eqnarray}
With $R^A_{\ \alpha} = g_{\alpha\ov\beta}R^{A\ov\beta}$ and
    $R^A_{\ \ov\alpha} = g_{\beta\ov\alpha}R^{A\beta}$
they satisfy the Killing equation
\begin{eqnarray} 
R^A_{\ \alpha,\ov\beta} + 
R^A_{\ \ov\beta,\alpha} \ \ = \ \ 0.\label{eq 25}
\end{eqnarray}
They also satisfy
\begin{eqnarray}
 R^A_{\ \alpha;\beta} = 0, \quad c.c., \label{eq 26}
\end{eqnarray}
from the holomorphic property. From (\ref{eq 25})
we may find real scalars 
$M^A(\phi,\bar\phi)$, called Killing potentials\cite{8}, such that
\begin{eqnarray}
R^A_{\ \alpha} = \ i M^A_{\ ,\alpha} , \quad \quad
R^A_{\ \ov \alpha} = - i M^A_{\ ,\ov \alpha}.   \label{eq 27}
\end{eqnarray}
These Killing potentials transform as the adjoint representation of the group $G$ by the Lie-variation
\begin{eqnarray}
{\cal L_{R^A}} M^B \equiv  R^{A\alpha} M^B_{\ ,\alpha} +
            R^{A\ov\alpha}M^B_{\ ,\ov\alpha} 
                     = f^{ABC}M^C,    \label{eq 28}
\end{eqnarray}
which may be written as
\begin{eqnarray}
R^{A\alpha}R^B_{\ \alpha} -
                      R^{B \alpha} R^A_{\ \alpha}
                   = if^{ABC}M^C, \label{eq 29}
\end{eqnarray}
by (\ref{eq 27}).
A manipulation of (\ref{eq 28}) with (\ref{eq 27}) leads us to write the Killing potentials in terms of the Killing vectors: 
\begin{eqnarray}
M^A = -{i\over {\cal N}_{adj}}  f^{ABC}R^{B\alpha}
R^{C\overline \beta} g_{\alpha \overline \beta}.   \label{eq 30}
\end{eqnarray}
Here we have used the normalization 
\begin{eqnarray}
f^{ABC}f^{ABD} = 2{\cal N}_{adj} \delta^{CD}. \nonumber
\end{eqnarray}

When the K\"ahler coset space $G/H$ is irreducible, there are relations between $M^A$ and the geometric quantities\cite{9}:
\begin{eqnarray}
{\rm scalar\ curvature}&:& \quad  \quad M^AM^A = R  \hspace{4cm}     \nonumber \\
& & \hspace{2.2cm} (= -R_{\alpha\ov\beta\gamma\ov\delta}g^{\alpha\ov\beta}g^{\gamma\ov\delta} = const.),
    \label{eq 31}     \\
{\rm metrics}&:& \quad\  M^A_{\ ,\alpha}M^A_{\ , \ov\beta} = g_{\alpha\ov\beta},
 \quad\  M^A_{\ ,\alpha}M^A_{\ , \beta} = 0,   \nonumber  \\
           & & \quad\ M^A_{\ ,\alpha,\ov\beta}M^A = -g_{\alpha\ov\beta}, \quad c.c., 
           \label{eq 32}\\
{\rm Riemann\ curvature}&:&  \ \
M^A_{\ ,\alpha,\ov\beta}M^A_{\ ,\gamma,\ov\delta} = - R_{\alpha\ov\beta\gamma\ov\delta}.  \label{eq 33}
\end{eqnarray}
In addition to these  we later need the following formula:
\begin{eqnarray}
\nabla_\eta R_{\alpha\ov\beta\gamma\ov\delta} = 0.\label{eq 34}
\end{eqnarray}
 To show the formula note that
\begin{eqnarray}
\nabla_\eta M^A_{\ ,\alpha,\ov\beta} =\nabla_\eta\nabla_{\ov\beta}M^A_{\ ,\alpha} =
R_{\eta\ov\beta\alpha}^{\ \ \ \ \sigma}  
M^A_{\ ,\sigma}\ ,  \nonumber
\end{eqnarray}
by (\ref{eq 26}) and (\ref{eq 27}). Using this and (\ref{eq 33}) we have 
\begin{eqnarray}
\nabla_\eta R_{\alpha\ov\beta\gamma\ov\delta} = -R_{\eta\ov\beta\alpha}^{\ \ \ \ \sigma}  
M^A_{\ ,\sigma}M^A_{\ ,\gamma,\ov\delta} -
R_{\eta\ov\delta\gamma}^{\ \ \ \ \sigma} M^A_{\ ,\alpha,\ov\beta}  
M^A_{\ ,\sigma}.  \nonumber
\end{eqnarray}
Each term of the {\it r.h.s.} vanishes because 
$$
M^A_{\ ,\sigma} M^A_{\ ,\gamma,\ov\delta} =
g_{\sigma\ov\lambda}R^{A\ov\lambda}R^A_{\ \ov\delta,\gamma} =
 g_{\sigma\ov\lambda}(R^{A\ov\lambda}R^A_{\ \ov\delta})_{ ,\gamma} = 0,
$$
by the holomorphic property of $R^{A\ov\lambda}$ and (\ref{eq 32}).

\hspace{2cm}

Equipped with these formulae let us study the $\star$-product for the irreducible K\"ahler coset space, defined by (\ref{eq 19}), and prove our claims (\ref{eq 0}) and (\ref{eq 00}). First of all we  solve eq. (\ref{eq 14}) for $r$. The first key observation is that due to (\ref{eq 34}) the iterative solution (\ref{eq 15}) does not get contribution from $\partial r$. Hence it turns out to take the form 
\begin{eqnarray}
r = \sum_{N=1}\sum_{ 1\le n \le 2N} \hbar^{2N-n}[\overbrace{R\otimes \cdots \otimes R}^N ]_{\alpha_1\alpha_2\cdots \alpha_n
\ov\beta_1\ov\beta_2\cdots \ov\beta_n}\theta^{\alpha_1}y^{\alpha_2}\cdots y^{\alpha_n}y^{\ov\beta_1}y^{\ov\beta_2}\cdots y^{\ov\beta_n},\nonumber \\
  -c.c.. \hspace{10cm}  \label{eq 35}
\end{eqnarray}
Here $[R\otimes \cdots \otimes R]_{\alpha_1\alpha_2\cdots \alpha_n\beta_1\beta_2\cdots \beta_n}$ is a symbolic notation for summing over all the types of the tensor product of $N$ Riemann curvatures, contracted $(2N-n)$ times by $g^{\alpha\ov\beta}$, but with no self-contraction like $R_{\alpha\ov\beta\gamma\ov\delta}g^{\gamma\ov\delta}$. For the sum we  understand appropriate coefficients obtained by iterating eq. (\ref{eq 14}). 
Next we deform $M^A$ according to (\ref{eq 16}) with the solution (\ref{eq 35}). That is, we solve eq. (\ref{eq 18}) for $a$ with the initial condition $a_0 = M^A$. Observe that due to (\ref{eq 34}) the iterative solution  consists of three types of quantities:
\begin{eqnarray}
{\rm Type}\ 1 :\hspace{14cm}    \nonumber \\
\hbar^{2N-n}[\overbrace{R\otimes \cdots \otimes R}^N \otimes M^A]_{\alpha_1\alpha_2\cdots \alpha_n
\ov\beta_1\ov\beta_2\cdots \ov\beta_n}y^{\alpha_1}y^{\alpha_2}\cdots y^{\alpha_n}y^{\ov\beta_1}y^{\ov\beta_2}\cdots y^{\ov\beta_n}, \hspace{1.5cm}  \nonumber   \\
    \quad\quad ({\rm degee}=4N), \hspace{5cm} \nonumber   \\
{\rm Type}\ 2  :\hspace{14cm}    \nonumber \\
\hbar^{2N-n}[\overbrace{R\otimes \cdots \otimes R}^N\otimes\nabla M^A]_{\alpha_1\alpha_2\cdots \alpha_{n+1}
\ov\beta_1\ov\beta_2\cdots \ov\beta_n}y^{\alpha_1}y^{\alpha_2}\cdots y^{\alpha_{n+1}}y^{\ov\beta_1}y^{\ov\beta_2}\cdots y^{\ov\beta_n} \hspace{1cm}  \nonumber \\
    +\  c.c., \hspace{4cm} ({\rm degee}=4N+1),   \hspace{4.5cm} \nonumber\\
{\rm Type}\ 3 :\hspace{14cm}    \nonumber \\
 \hbar^{2N-n}[\overbrace{R\otimes \cdots \otimes R}^N\otimes\ov\nabla\nabla M^A]_{\alpha_1\alpha_2\cdots \alpha_{n+1}
\ov\beta_1\ov\beta_2\cdots \ov\beta_{n+1}}y^{\alpha_1}y^{\alpha_2}\cdots y^{\alpha_{n+1}}y^{\ov\beta_1}y^{\ov\beta_2}\cdots y^{\ov\beta_{n+1}}, 
 \nonumber  \\
    \quad\quad ({\rm degee}=4N+2).   \hspace{4.5cm} \nonumber
\end{eqnarray}
The quantities with higher covariant derivatives of $M^A$ always reduce to the Type 2,  or zero,
for instance
\begin{eqnarray}
\nabla_\alpha\nabla_{\ov\beta}\nabla_{\gamma}M^A = 
 R_{\alpha\ov\beta\gamma}^{\ \ \ \ \delta}\nabla_\delta M^A, \quad\quad \nabla_\alpha\nabla_{\beta}\nabla_{\gamma}M^A = 0,  \nonumber
\end{eqnarray}
by (\ref{eq 26}). 
By using the deformed Killing potentials  we calculate the {\it l.h.s} of (\ref{eq 0}), denoted by $[M^A,M^B]_\star$. All tensor indices in the $\circ$ product are contracted by $g_{\alpha\ov\beta}$. There is no contribution from an even number of  contractions. So it suffices to calculate the $\circ$ product of the quantities of Type 2:
\begin{eqnarray}
[M^A,M^B]_\star = \hbar^{2(N+N') + 1}\sum_{N,N'}[\overbrace{R\otimes \cdots \otimes R}^{N+N'}]_\alpha^{\ \beta}
(R^{A\alpha}R^B_{\ \beta} - R^{B\alpha}R^A_{\ \beta}), \nonumber
\end{eqnarray}
by (\ref{eq 27}).
If we have a formula such that
\begin{eqnarray}
[\overbrace{R\otimes \cdots \otimes R}^N]_\alpha^{\ \beta} = const. \delta_\alpha^{\ \beta},  \label{eq 36}
\end{eqnarray}
then the claim (\ref{eq 0}) is shown by (\ref{eq 29}).
 If given the formula (\ref{eq 36}), we can also prove the claim (\ref{eq 00}). This time an odd number of contractions do not contribute to $M^A\star M^A$. Therefore we have 
\begin{eqnarray}
M^A\star M^A &=&  \hbar^{2(N+N')}\sum_{N,N'}[\overbrace{R\otimes \cdots \otimes R}^N\otimes M^A\otimes M^A]_\alpha^{\ \alpha}  
       \nonumber \\
    & &  \  +\  \hbar^{2(N+N')+2}\sum_{N,N'}[\overbrace{R\otimes \cdots \otimes R}^{N+N'} \otimes \ov\nabla\nabla M^A
    \otimes \ov\nabla\nabla M^A]_\alpha^{\ \alpha}. \nonumber
\end{eqnarray}
By (\ref{eq 31}) and (\ref{eq 33}) it reduces to 
\begin{eqnarray}
M^A\star M^A &= &  \hbar^{2(N+N')}\sum_{N,N'}[\overbrace{R\otimes \cdots \otimes R}^{N+N'}]_\alpha^{\ \alpha}\cdot const. \hspace{3cm}   \nonumber \\
   & &    \ -\  \hbar^{2(N+N')+2}\sum_{N,N'}[\overbrace{R\otimes \cdots \otimes R}^{N+N'+1}]_\alpha^{\ \alpha}. \nonumber
\end{eqnarray}
So the claim (\ref{eq 00}) is proved by the formula (\ref{eq 36}).

\hspace{2cm}

Let us now show the formula. To this end we remind of the method for constructing the K\"ahler coset space $G/H$\cite{10}. We consider the irreducible case. Then the group $G$ have generators $T^A =\{ X_\alpha,\ov X^\alpha, H^i,Y \}$   which satisfy the Lie-algebra
\begin{eqnarray}
[ X_\alpha, \ov X^\beta ] &=&  p(\Gamma^i)_\alpha^\beta H^i +
 q \delta_\alpha^\beta Y,  \quad\quad  [ X_\alpha, X_\beta ] \ =\  0,  \nonumber \\
 \quad [ X_\alpha, H^i ] &=& (\Gamma^i)_\alpha^\beta X_\beta,\quad\quad [X_\alpha, Y]\ =\  X_\alpha, \quad c.c.,   \label{eq 36'}
\end{eqnarray}
with some constants $p$ and $q$ depending on the representation of $G$.
Here $X_\alpha$ and $\ov X^\alpha$ are coset generators. In the method of ref. \cite{10} the local coordinates of $G/H$ are denoted by $z_\alpha$ and $\ov z^\alpha$, where upper or lower indices stand for complex conjugation. Therefore raising or lowering tensor indices in the foregoing discussions is done by writing the metrics $g_\alpha^{\ \beta}$ or $(g^{-1})_\alpha^{\ \beta}$ explicitly. Simple algebra gives 
\begin{eqnarray}
[X_\alpha, [X_\gamma, \ov X^\beta ] ] &=& \{p (\Gamma^i)_\alpha^\beta(\Gamma^i)_\gamma^\delta + q \delta_\alpha^\beta \delta_\gamma^\delta)\} X_\delta \nonumber \\
 & \equiv & M_{\alpha\gamma}^{\beta\delta} X_\delta  \label{eq 37}
\end{eqnarray}
The Killing vectors  $R^{A\alpha}\ (R^{A\ov \alpha})$ are nothing but non-linear realizations of the Lie-algebra (\ref{eq 36'}) on $z^\alpha\ (\ov z^\alpha) $:
\begin{eqnarray}
R^A_{\ \alpha} \equiv -i[T^A, z_\alpha], \quad c.c..  \nonumber
\end{eqnarray}
Then the Lie-algebra (\ref{eq 24}) is equivalent to the Jacobi identities
\begin{eqnarray}
[T^A,[T^B, z_\alpha]] - [T^B,[T^A,z_\alpha]] = [[T^A,T^B],z_\alpha].   \nonumber
\end{eqnarray}
By examining these Jacobi identities the Killing vectors were shown to take the forms \begin{eqnarray}
R^\gamma_{\ \alpha} &=& i\delta^\gamma_\alpha, \quad\quad 
R_{\gamma\alpha} \ =\ {i\over 2}M_{\alpha\gamma}^{\beta\delta} z_\beta z_\delta,
 \nonumber  \\
R^i_{\ \alpha} &=&  i(\Gamma^i)_\alpha^\beta z_\beta, \quad\quad
R_\alpha \ =\ i z_\alpha,   \nonumber
\end{eqnarray}
owing to the formulae
\begin{eqnarray}
M_{\alpha\gamma}^{\beta\delta} &=& M_{\gamma\alpha}^{\beta\delta} \ =\  M_{\alpha\gamma}^{\delta\beta} \ = \  M_{\gamma\alpha}^{\delta\beta},   \nonumber \\
M_{\rho\sigma}^{\lambda(\alpha}M_{\tau\lambda}^{\beta\gamma)} &=&
M_{\sigma\tau}^{\lambda(\alpha}M_{\rho\lambda}^{\beta\gamma)} \ =\ 
M_{\tau\rho}^{\lambda(\alpha}M_{\sigma\lambda}^{\beta\gamma)},    \nonumber
\end{eqnarray}
where the round brackets denote complete symmetrization over all three indices enclosed. The quantity $M_{\alpha\gamma}^{\beta\delta}$ defined by (\ref{eq 37}) is a building block for constructing the manifold. The Riemann curvature takes the form\cite{10} 
$$
R_{\alpha\ \gamma}^{\ \beta \ \delta} = M_{\alpha\gamma}^{\kappa\lambda}
g_{\kappa}^{\ \beta} g_\lambda^{\ \delta}.
$$
By using this  form of the Riemann curvature  we obtain 
\begin{eqnarray}
[\overbrace{R\otimes \cdots \otimes R}^N]_{\ \alpha_1\alpha_2\cdots \alpha_n}^{\ \ \gamma_1\gamma_2\cdots \gamma_n}
(g^{-1})_{\gamma_1}^{\ \beta_1}(g^{-1})_{\gamma_2}^{\ \beta_2}\cdots (g^{-1})_{\gamma_n}^{\ \beta_n}  \hspace{2cm} \nonumber \\
 \ = \  [[ \overbrace{M\otimes \cdots \otimes M}^N ]]_{\ \alpha_1\alpha_2\cdots \alpha_n}^{\ \beta_1\beta_2\cdots \beta_n}.  \label{eq 38}         
\end{eqnarray}
Here the symbol $[[\ ]]$ stands for the same contraction as $[\ ]$, but 
 by the Kronecker delta $\delta_\alpha^\beta$, in place of the metric $(g^{-1})_\alpha^{\ \beta}$. In the case of $n=1$ we calculate the $r.h.s$, using 
 $M_{\alpha\gamma}^{\beta\delta}$, defined by (\ref{eq 37}), $\Gamma^i\Gamma^i = const.$ and the Lie-algebra of $\Gamma^i$,
 to show the formula (\ref{eq 36}).

\hspace{2cm}

Finally we apply the formalism to the case of $CP^1$ as a demonstration. For $CP^1$ with the line element given by 
$$
ds^2 = {1\over (1+{1\over 2}|z|^2 )^2} dz d\ov z,
$$
we have the non-trivial components of the Riemann curvature 
$$
R_{z\ov z z\ov z} = -{1 \over (1+{1\over 2}|z|^2)^4} ,
$$
and the Killing potentials $M^A$:
$$
M^+ = {z\over 1+{1\over 2}|z|^2 },\quad M^- = {\ov z \over 1+{1\over 2}|z|^2 }, \quad 
M^0 = {1-{1\over 2}|z|^2 \over 1+{1\over 2}|z|^2 }.
$$
Then the solution (\ref{eq 15}) is given by 
$$
r \ =\  -{i\over 4}{1\over (1+{1\over 2}|z|^2)^4} |y|^2(y\ov \theta - \ov y \theta) \ +\
 {i\hbar^2 \over 64}{1\over (1+{1\over 2}|z|^2)^2}(y\ov \theta - \ov y \theta) + \cdots.
$$
With this $r$ we find the deformed Killing potentials $\hat M^A$:
\begin{eqnarray}
\hat M^A \ =\ M^A &+& 
[\{(1-{13 \over 192}\hbar^2)y \ +\  {5\over 12}{ 1\over (1+{1\over 2}|z|^2)^2} y |y|^2  \cdots \ \}M^A_{\ ,z}  + c.c. ]  \nonumber  \\
    &+& [(1-{1\over 8}\hbar^2 )|y|^2 \ + \ 
{1\over 3}{ 1\over (1+{1\over 2}|z|^2)^2} |y|^4 \cdots \ ] M^A_{\ ,z,\ov z}.  \nonumber 
\end{eqnarray}
The fuzzy algebrae (\ref{eq 0}) and (\ref{eq 00}) take the forms 
\begin{eqnarray}
[M^A,M^B]_\star &=& -i[\ \hbar  -  {2\over 9}\hbar^3 \ +\ O(\hbar^5)\ ]\  
\varepsilon^{ABC}M^C, \nonumber \\
     & &       \nonumber \\
M^A\star M^A &=& \ 1  -  {1\over 4} \hbar^2  +  {13\over 144} \hbar^4 \ +\  O(\hbar^6\ ). 
\nonumber
\end{eqnarray}
The coefficients differ from those obtained in ref. \cite{5}.  Using the the spherical coordinates, they found that $c_n = 0$ for $n \ge 3$. 
Apparently the coefficients of the fuzzy algebrae  depend on the reparametrization of the coset space. 

\hspace{2cm}

In this paper we have shown the fuzzy algebrae on the irreducible K\"ahler coset space by  Fedosov's $\star$ product. It is desirable to extend the study to the reducible K\"ahler coset space by using the formalism in ref. \cite{11}. 
The calculations of the deformed K\"ahler potentials $\hat M^A$ were rather 
complicated even in the case of $CP^1$. They would be simplified by using a modified version of the $\star$ product, which was discussed for the K\"ahler manifold in \cite{12}. 
It is interesting to develop a diagramatical method which enables us to calculate the $\star$ product systematically. The study in this direction is on the course.

\vspace{2cm}
\noindent
{\Large\bf Acknowledgements}

The work of T.M. was supported in part by JSPS Research Fellowships for Young Scientists.

\hspace{3cm}


\begin{thebibliography}{99}

\bibitem{1} N. Seiberg and E. Witten, ``String Theory and Noncommutative Geometry",
JHEP {\bf 9909}(1999)032, hep-th/9908142.
\bibitem{1'}
N. Nekrasov and A. Schwarz, ``Instantons on Noncommutative $R^4$, and $(2,0)$ Superconformal Six Dimensional Theory", Comm. Math. Phys. {\bf 198}(1998)689, hep-th/9802068;

S. Minwalla, M. Van Raamsdonk and N. Seiberg, ``Noncommutative Perturbative Dynamics", JHEP {\bf 0002}(2000)020, hep-th/9912072;

D. Gross and N. Nekrasov, ``Monopoles and Strings in Noncommutative Gauge Theory", 
JHEP {\bf 0007}(2000)034, hep-th/0005204;

 R. Gopakumar, S. Minwalla and A. Strominger, ``Noncommutative Solitons", JHEP 
 {\bf 0005}(2000)020, hep-th/0003160;

J.A. Harvey, P. Kraus, F. Larsen and E.J. Martinec, ``D-branes and Strings as Non-commutative Solitons", JHEP {\bf 0007}(2000)042, hep-th/0005031.
\bibitem{1''} L. Cornalba and R. Schiappa, ``Nonassociative Star Product Defformations for $D$-Brane Worldvolumes in Curved Backgrounds", hep-th/0101219.
\bibitem{2} N. Reshetikhin and L.A. Takhtajan, ``Deformation Quantization of K\"ahler manifolds", math.QA/9907171.

\bibitem{3} A.S. Cattaneo and G. Felder, ``A Path integral Approach to the Kontsevitch Quantization Formula", Comm. Math. Phys. {\bf 212}(2000)591, math.QA/9902090.
\bibitem{4} M. Kontsevitch, ``Deformation Quantization of Poisson Manifolds, I", q-alg/9709040.
\bibitem{5} I. Kishimoto, ``Fuzzy Sphere and Hyperbolic Space from Deformation Quantization", JHEP {\bf 0103}(2001)025, hep-th/0103018.
\bibitem{6} B.V. Fedosov, ``A Simple Geometrical Construction of Deformation Quantization", J. Diff. Geom. {\bf 40}(1994)213.

B.V. Fedosov, ``Deformation Quantization and Index theory", Berlin, Germany : Akademie-Verl. (1996) (Mathematical Topics : {\bf 9}).
\bibitem{7} J. Hoppe, ``Diffeomorphism Groups, Quantization and $SU(\infty )$", Int. J. Mod. Phys. {\bf A4}(1989)5235;

J. Madore, ``The Fuzzy Sphere", Class. Quant. Grav. {\bf 9}(1992)69;

U. Carow-Watamura and S. Watamura, ``Noncommutative Geometry and Gauged Theory on Fuzzy Sphere", Comm. Math. Phys. {\bf 212}(2000)395, hep-th/9801195;

A. Alekseev, A. Recknagel and V. Schomerus, ``Open Strings and Non-commutative Geometry of Branes on Group Manifolds", Mod. Phys. Lett. {\bf A16}(2001)325, hep-th/0104054.
\bibitem{8} J. Bagger and E. Witten, ``The Gauge Invariant Supersymmetric Nonlinear Sigma Model", Phys. Lett. {\bf 118B}(1982)103.
\bibitem{9} S. Aoyama, ``A New Visualization of the Gauged Supersymmetric $\sigma$-model by Killing Potentials", Z. Phys. C. {\bf 32}(1986)113.
\bibitem{10} J. Achiman, S. Aoyama and J.W. van Holten, ``The Non-linear Supersymmetric $\sigma$-model on $E_6/SO(10)\otimes U(1)$", Phys. Lett. {\bf 141B}(1984)64; ``Gauged Supersymmetric $\sigma$-models and $E_6/SO(10)\otimes U(1)$", Nucl. Phys. {\bf B258}(1985)179.
\bibitem{11} S. Aoyama, ``The Four-fermi Coupling of the Supersymmetric Non-linear $\sigma$-model on $G/S\otimes\{U(1)\}^k$ ", Nucl. Phys. {\bf B578}(2000)449, hep-th/0001160.
\bibitem{12} M. Bordemann and S. Waldmann, ``A Fedosov Star Product of Wick Type for K\"ahler Manifolds", q-alg/9605012.


.



\end{thebibliography}
\end{document}